\documentclass[]{spie}  

\usepackage{amsmath,amsfonts,amssymb}
\usepackage{graphicx}
\usepackage[colorlinks=true, allcolors=blue]{hyperref}

\newcommand{\norm}[1]{\left\lVert#1\right\rVert}

\usepackage{soul}

\title{Automatic evaluation of human oocyte developmental potential from microscopy images}

\author[a]{Denis Baručić}
\author[a]{Jan Kybic}
\author[b]{Olga Teplá}
\author[b]{Zinovij Topurko}
\author[c]{Irena Kratochvílová}
\affil[a]{Department of Cybernetics, Faculty of Electrical Engineering, Czech Technical University in Prague, Czech Republic}
\affil[b]{Department of Obstetrics and Gynecology, The First Faculty of Medicine and General Teaching Hospital, Czech Republic}
\affil[c]{Institute of Physics of the Czech Academy of Sciences, Czech Republic}

\begin{document} 
\maketitle

\begin{abstract}
Infertility is becoming an issue for an increasing number of couples. The most
common solution, in vitro fertilization, requires embryologists to carefully
examine light microscopy images of human oocytes to determine their
developmental potential. We propose an automatic system to improve the speed,
repeatability, and accuracy of this process. We first localize individual
oocytes and identify their principal components using CNN (U-Net) segmentation.
Next, we calculate several descriptors based on geometry and texture. The final step
is an SVM classifier. Both the segmentation and classification training is
based on expert annotations. The presented approach leads to a classification
accuracy of $70\%$.
\end{abstract}

\keywords{human oocytes, fertilization, microscopy, classification, segmentation}

\section{Introduction}
\label{sec:intro}

Infertility has been an issue for several years and is expected to grow further. Nowadays, the most common solution for an infertile couple is in vitro
fertilization (IVF). One of the crucial steps is choosing the best oocytes to
be fertilized, since, for practical, ethical, and legal reasons, it is not
feasible to fertilize more than a few of them --- and even fewer can be
implanted.

The situation is easier when a patient's own oocytes are used and are,
therefore, more readily available. In this case, the embryologists attempt to
fertilize almost all collected oocytes that are not apparently damaged. The
knowledge gained from these attempts can be used later to determine the a priori
developmental potential of newly collected oocytes, reducing both the cost and
the failure rate of the IVF. This is especially important when the oocytes come
from a~donor. In this case, each oocyte is very valuable, and it is important
to determine its developmental potential reliably. Currently, this is performed
by an expert who carefully examines the oocytes under a microscope. The
selection process requires extensive experience, is time-consuming, and is done
outside the optimal environment, so it is desirable to shorten it as
much as possible.

In this work, we aim to replace this subjective process with an automatic
evaluation of the developmental ability of oocytes from digitized light
microscopy images to improve its speed, repeatability, and accuracy. We present
a proof-of-concept solution on a~small dataset, showing the viability of this
approach.

\subsection{Previous work}

To the best of our knowledge, the task of \emph{fully} automatic oocyte developmental
potential assessment has not been addressed before. The closest work to ours is
Manna et~al.~\cite{manna2013}, who feed LBP texture features to an ensemble of
shallow neural networks and try to predict whether an oocyte or embryo (i.e.,
fertilized oocyte) will lead to birth. However, their approach requires manual segmentation, and obtaining the ground truth data takes much more time. Since multiple embryos
are usually implanted, the outcome is only certain in the relatively few cases (namely 12\cite{manna2013}) when all or none of the implanted embryos lead to birth. Furthermore, many
embryos are not implanted but frozen, and the pregnancy outcome might not be
known for years or not at all. For this reason, here we have chosen a different classification target and decided to predict an
embryo's ability to start the correct development, which can be determined
relatively quickly and for all embryos.

Recently, deep learning has been used for oocyte segmentation\cite{firuzinia2021,targosz2021}. Other works analyzed microscopic images of embryos\cite{khosravi2019, raudonis2019}. An extensive database of $50\,000$ embryo images allowed to train a model based on deep learning to classify the embryos into three
quality classes\cite{khosravi2019}. Raudonis et~al.\cite{raudonis2019} described a method for analyzing time-lapse sequences of embryo images.

Viswanath et~al.\cite{viswanath2016} classified swine cumulus oocyte complexes (i.e., oocytes with the cumulus cells, unlike in our data). The
authors examined the number of cumulus cell layers and the homogeneity of the cytoplasm. A semi-automatic (snake) method was used for segmentation and random forests for the classification of oocytes. Pure texture
analysis turned out to be useful for cytoplasm clustering{\cite{basile2010}}
or cytoplasm segmentation{\cite{caponetti2011}}.

\subsection{Proposed approach}

\begin{figure}[tb]
\centering
\includegraphics[width=55mm]{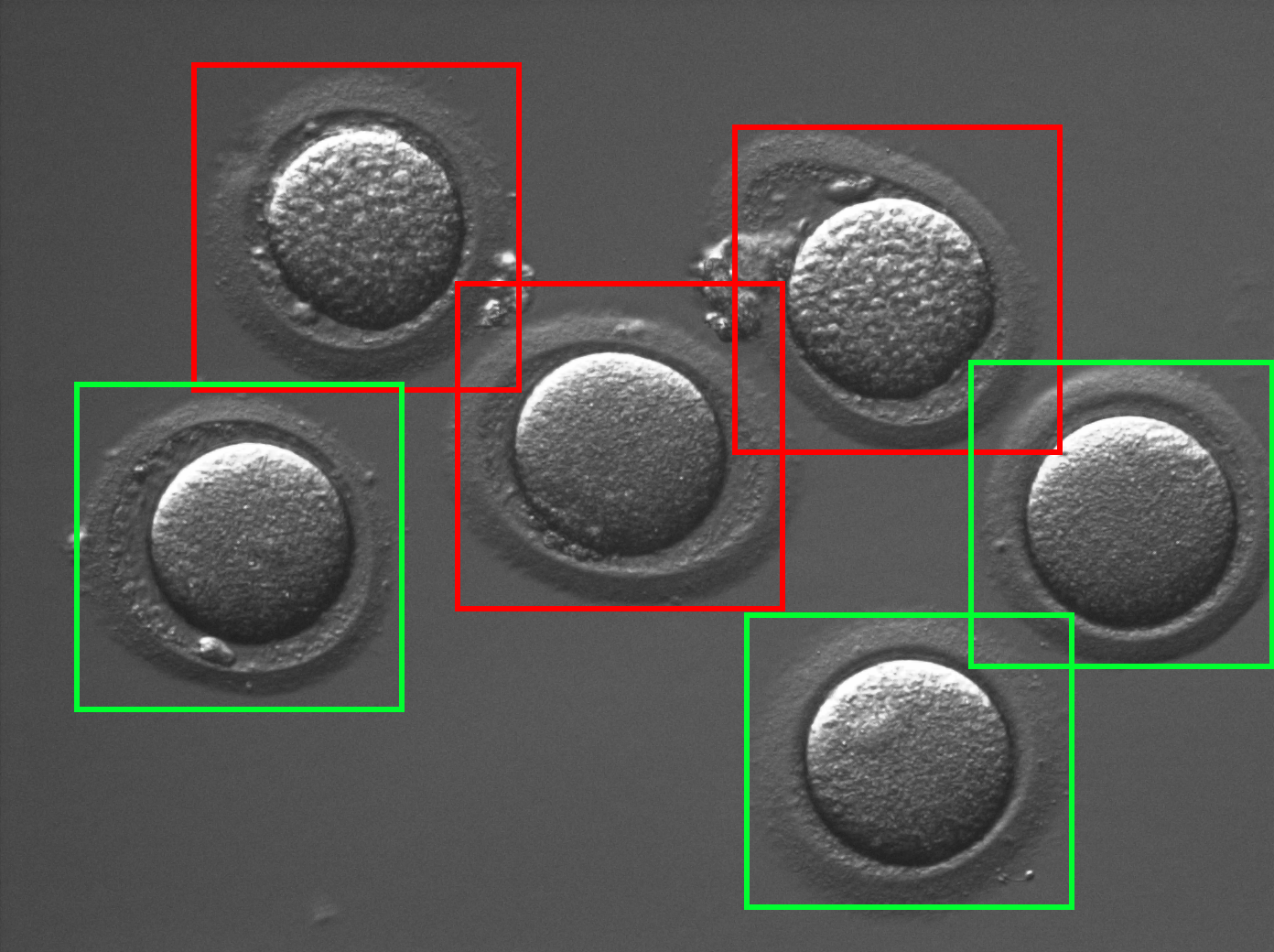}
\includegraphics[width=55mm]{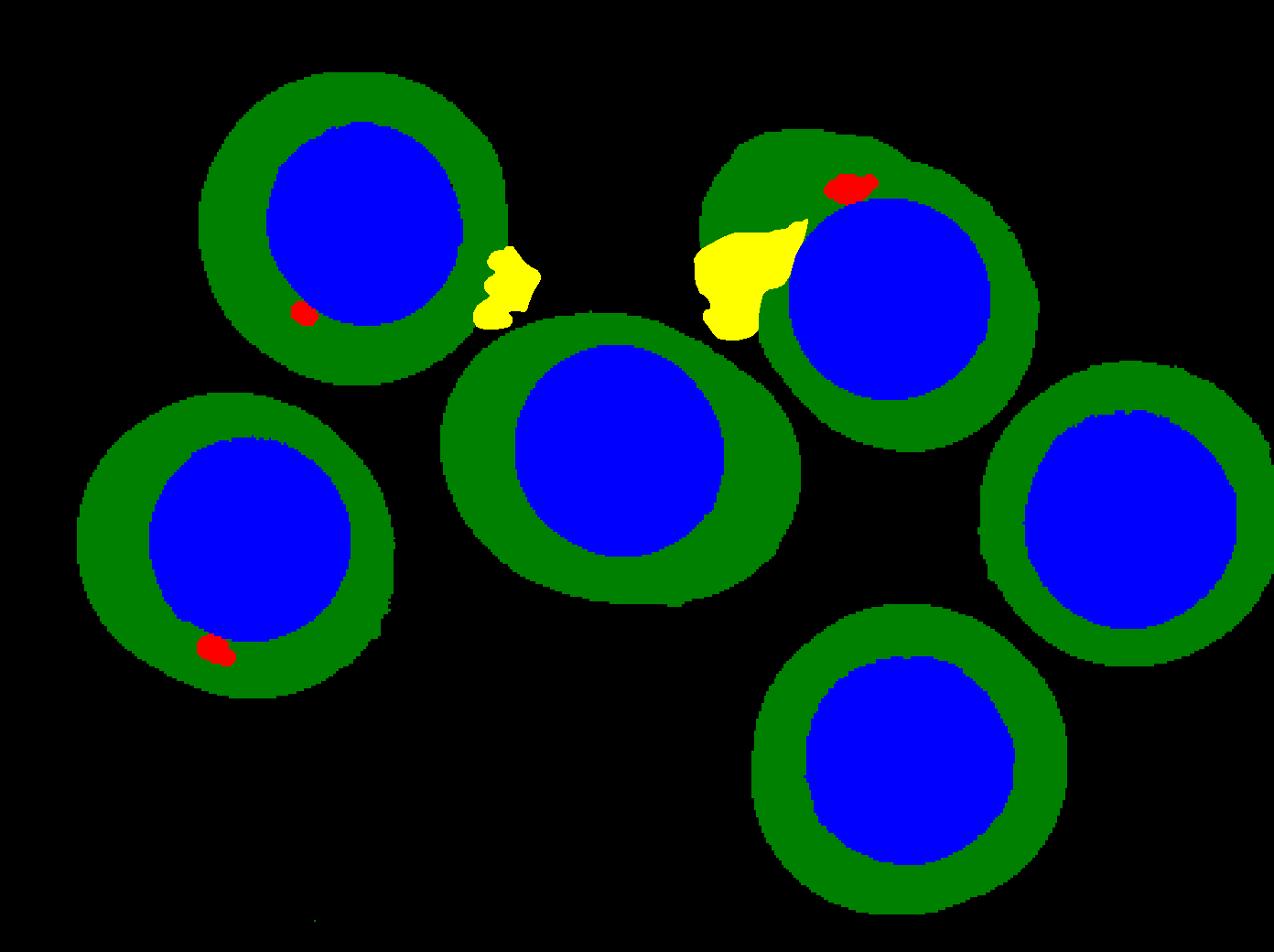}
\caption{Example of an input image and the corresponding expert segmentation.
 The green and red frames denote viable and nonviable oocytes, respectively (left image). Classes \textit{background}, \textit{cytoplasm}, \textit{zona pellucida},
\textit{polar body}, and \textit{cumulus cells} are denoted in black, blue,
green, red, and yellow, respectively (right image).}
\label{fig:image-sample}
\end{figure}

This paper describes a fully automatic approach to classify oocytes in light
microscopy images (see Fig.~\ref{fig:image-sample}) into two categories, viable
and nonviable, where viable oocytes have a~good potential of becoming
well-developed embryos. We learn from subjective expert annotations of
individual oocytes.  Our approach 
consists of five consecutive stages --- localization, extraction of individual oocytes,
segmentation, feature extraction, and classification.

\section{Data}
\label{sec:data}

Our anonymized dataset consists of 34 grayscale images of groups of oocytes
after cumulus cell denudation. Each image contains $1\sim 7$ oocytes. The images of $1392 \times 1040$ pixels were
acquired using Nikon Diaphot 300 inverted microscope, Eppendorf (Hamburg,
Germany) micromanipulation system equipped with a~thermoplate (Tokai Hit,
Japan). The acquisition of the images was approved by the Ethics Committee of the General University Hospital, Prague, Czech Republic, on October 12, 2018, reference number 79/17.

The ground truth (GT) segmentations (see Fig.~\ref{fig:image-sample}) were
created using the GIMP graphical editor. We considered four classes: \textit{background}, \textit{cytoplasm}, \textit{zona pellucida}, \textit{polar body}, and \textit{cumulus cells}.  The individual oocytes were classified as viable or nonviable
by the embryology expert (OT). Skipping incompletely visible oocytes yielded 50 viable and 53
nonviable oocytes. Furthermore, from
observations of the development, we learned the true number of viable oocytes $y_j$ in
each image $j$. This information is used to check the expert annotations in Sec.~{\ref{sec:expert-accuracy}}.

\section{Method}
\label{sec:method}

\begin{figure}[tb]
\includegraphics[width=\linewidth]{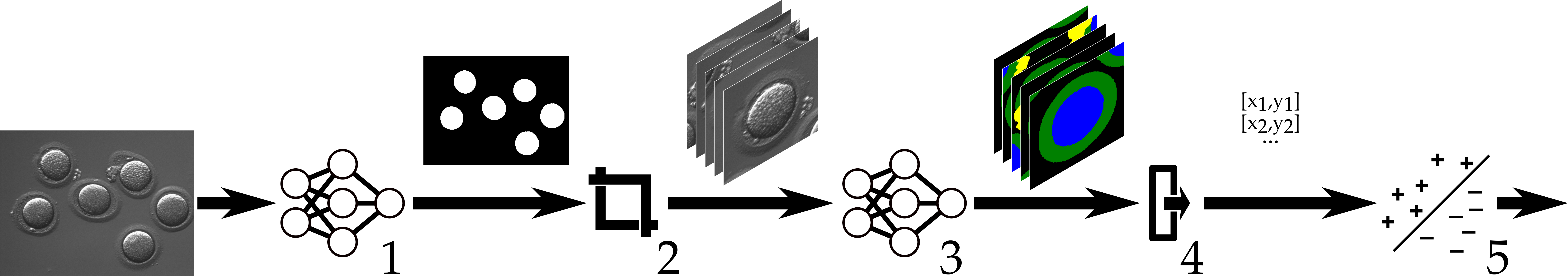}
\caption{The classification pipeline. (1)~Binary segmentation. (2)~ROI
extraction. (3)~Five class segmentation. (4)~Feature extraction. (5)~Classification.}
\label{fig:pipeline}
\end{figure}

Our method consists of five stages, explained in the following subsections. Fig.~\ref{fig:pipeline} gives a~high-level overview.

\subsection{Oocyte localization}
\label{sec:localization}

We first perform a~binary segmentation of \textit{cytoplasm} versus
the remaining classes because the cytoplasm is clearly distinguishable in the
images.  We use a~U-Net \cite{ronneberger2015} CNN with MobileNetV2
\cite{sandler2018} as the encoder, the standard mirroring decoder, and
a~pixel-wise softmax final layer. MobileNetV2 is a~fast architecture with a
relatively low number of parameters. The extra speed is desirable when the tool
is deployed in a production environment. The network was trained for 500 epochs
until convergence. We use the Dice loss function in combination with the ADAM optimizer
(learning rate $10^{-4}$). To prevent overfitting, multiple data augmentation
methods (shifting and rotation, contrast and brightness adjustments, and
blurring) are applied to the training images during training.

\begin{figure}[tb]
\centering
\begin{minipage}{55mm}
\includegraphics[width=\linewidth]{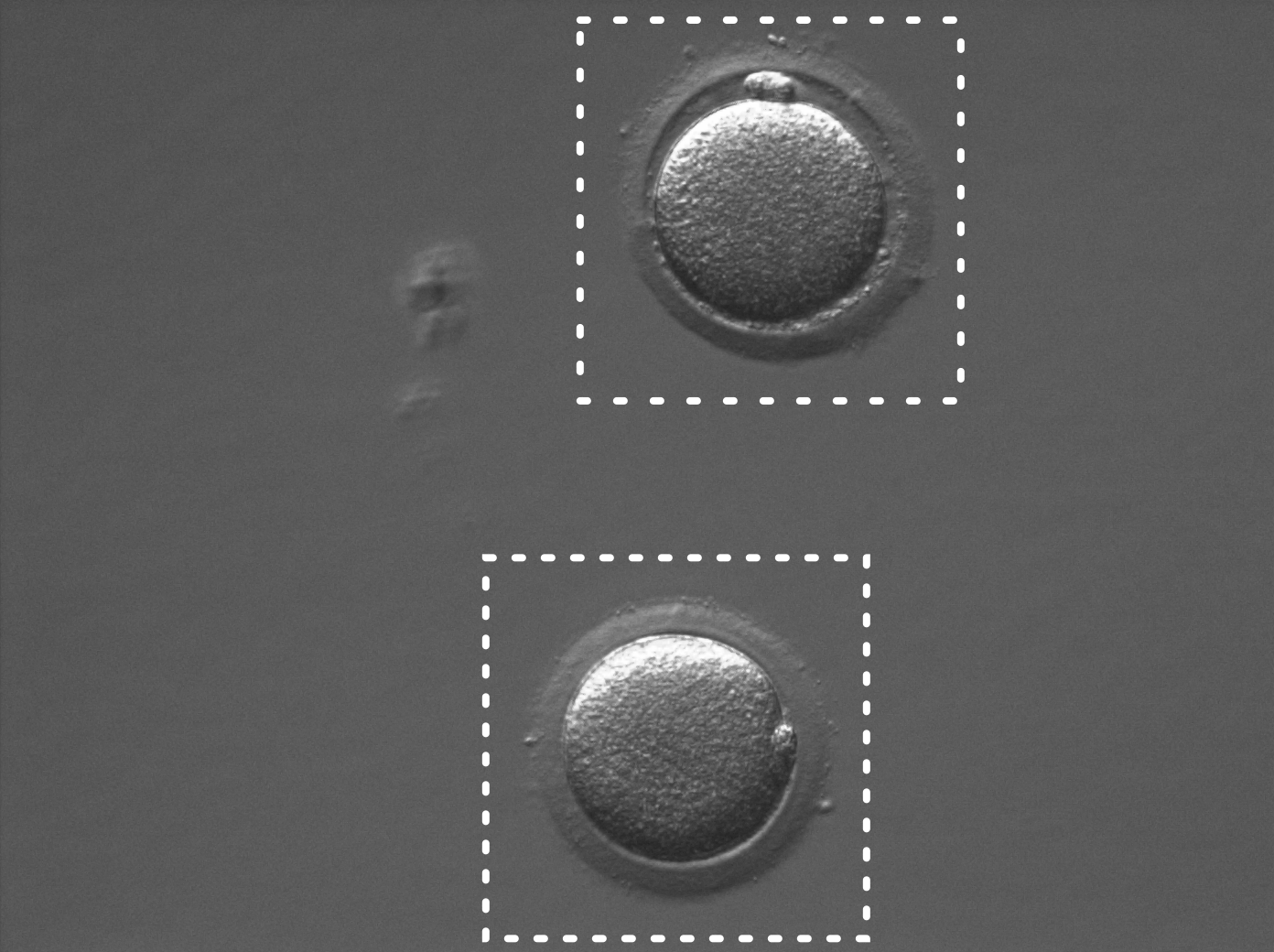}
\end{minipage}
\begin{minipage}{20mm}
\includegraphics[width=\linewidth]{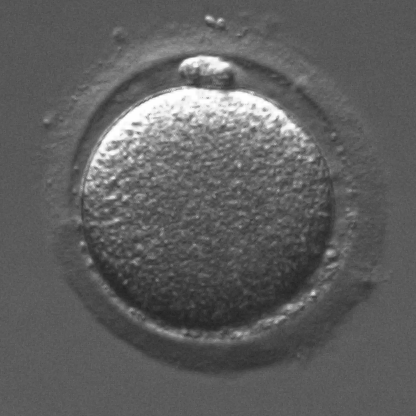}\vspace{.5mm}
\includegraphics[width=\linewidth]{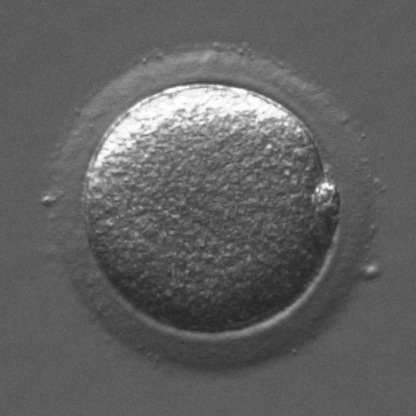}
\end{minipage}
\caption{Example of an input image (left) and two ROIs extracted from the
image.}
\label{fig:extraction}
\end{figure}

Connected foreground components smaller than $10^4$ pixels are suppressed.
The threshold was picked so that it does not rule out any viable oocytes, the
area of which is always more than $4\cdot 10^4$ pixels.  Finally, regions of interest (ROI)
of size $416 \times 416$ are extracted from around the centers of gravity of
the remaining foreground components (see Fig.~\ref{fig:extraction}).

\subsection{Oocyte segmentation}
\label{sec:segmentation}

\begin{figure}[t]
\centering
\includegraphics[width=25mm]{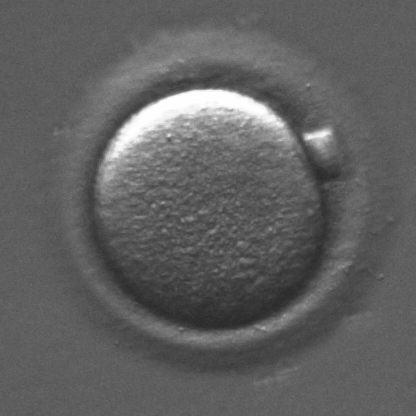}
\includegraphics[width=25mm]{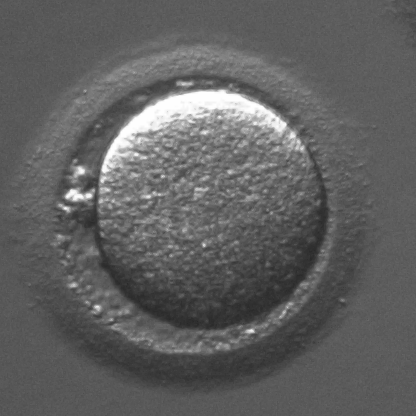}
\includegraphics[width=25mm]{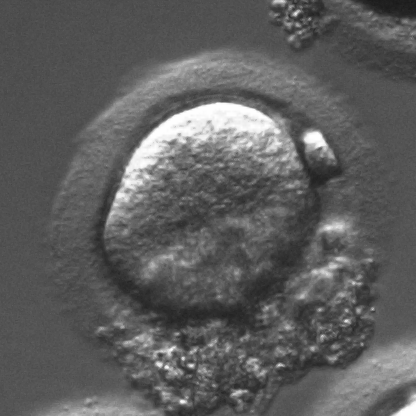}
\includegraphics[width=25mm]{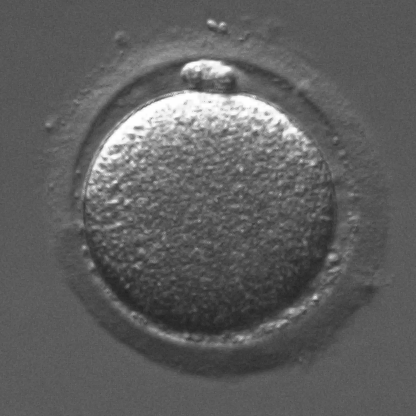}

\includegraphics[width=25mm]{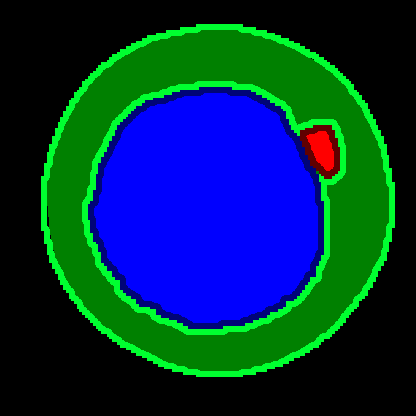}
\includegraphics[width=25mm]{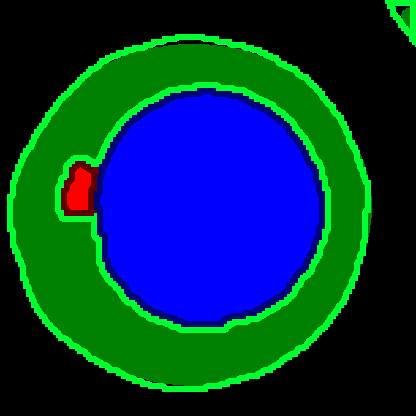}
\includegraphics[width=25mm]{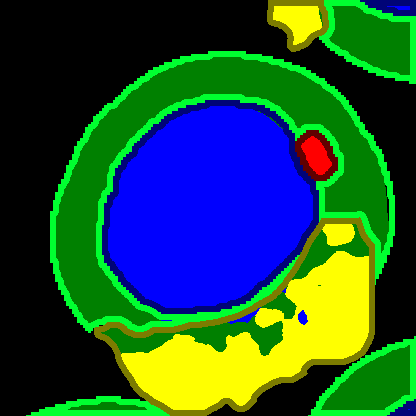}
\includegraphics[width=25mm]{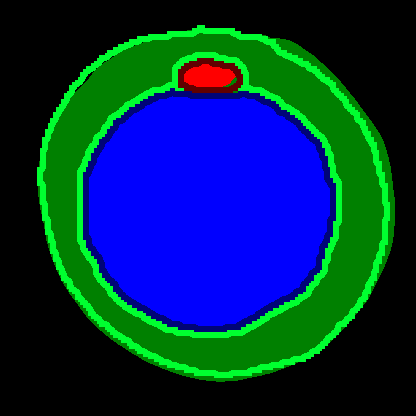}
\caption{Four extracted ROIs in the top row and the corresponding predicted
segmentations (black, green, blue, red and yellow denote background, zona pellucida, cytoplasm, polar body and cumules cells, respectively) and expert segmentations (outlined).}
\label{fig:segmentation}
\end{figure}

Once the ROIs are extracted, they are segmented into the five classes
described in Sec.~\ref{sec:data} using another CNN. Since the \textit{polar body} or
\textit{cumulus cells} classes are challenging to segment, we use a U-Net with the powerful
ResNet50 architecture~\cite{he2016} as the encoder, trained for 600 epochs,
which was sufficient for convergence. The rest of the procedure is identical to
Sec.~\ref{sec:localization}. Example segmentations are shown in
Fig.~\ref{fig:segmentation}.

\subsection{Feature extraction}

Given the small number of available training images, we could not use deep learning for feature extraction. Instead, using the segmentation from the previous section (ROI) we
compute for each oocyte the 24 features described below.

First, to handle the case where the ROI contains parts of several oocytes, we
keep only the largest \textit{cytoplasm} and \textit{zona pellucida} components.  We also suppress
\textit{polar body} components smaller than 500 pixels (for a~bad oocyte, it is
possible to have multiple \textit{polar body} components, so we cannot just keep the
largest).

\begin{figure}[tb]
\centering
\includegraphics[width=25mm]{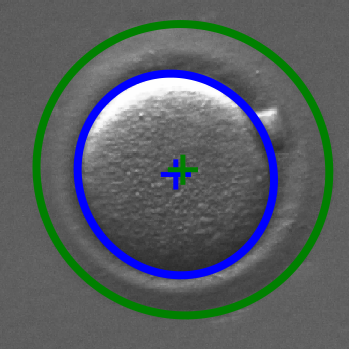}
\includegraphics[width=25mm]{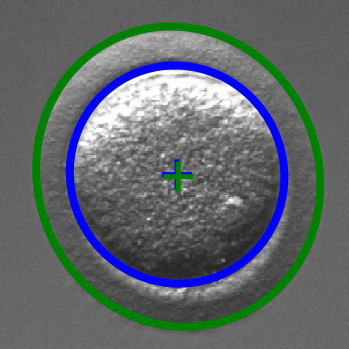}
\includegraphics[width=25mm]{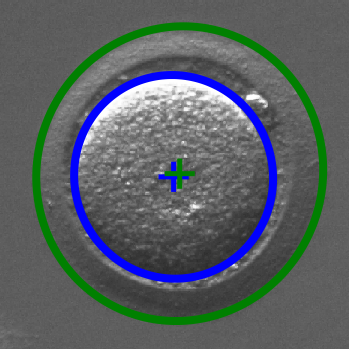}
\caption{Fitted ellipses for \textit{cytoplasm} (blue) and \textit{zona pellucida} (green).}
\label{fig:ellipses}
\end{figure}

Ellipses are fitted to the boundary of the \textit{cytoplasm} class and to the outer
boundary of the \textit{zona pellucida} class by least squares fitting of the boundary pixels (see
Fig.~\ref{fig:ellipses}). We calculate the following features based on the
cytoplasm ellipse:
\begin{itemize}
\item {mean axis} $\mu_{c}=\frac{a_{c}+b_{c}}{2}$,
\item {eccentricity}
$e_{c}=\sqrt{1-\frac{a_{c}^2}{b_{c}^2}}, a_{c} \geq b_{c}$,
\item {compactness} $\gamma_{c}=\frac{a_{c}b_{c}\pi}{S_{c}}$,
\end{itemize}
where $a_{c}, b_{c}$ are the estimated ellipse semi-axes, and $S_{c}$ is the area of
the \textit{cytoplasm} component. The features $\mu_{z}, e_{z}, \gamma_{z}$ are
calculated similarly for the \textit{zona pellucida} class.  We also define the misalignment $m=\norm{\textbf{c}_{c}-\textbf{c}_{z}}$, the Euclidean distance
between the ellipse centers, and the ratio of the cytoplasm and zona pellucida
areas\cite{basile2010}, $r=\frac{S_{c}}{S_{z}}$.

Regarding polar bodies, two features are used: the number of connected
components, $n_{pb}$, and the total area, $S_{pb}$.  The presence of cumulus
cells is not related to the oocyte quality but may influence the other
features, hence we also calculate the total area of cumulus cells $S_{cc}$,
which completes the 11 geometrical features.

The remaining 13 features describe the texture of the cytoplasm\cite{manna2013}. The first 10 texture features are calculated
efficiently from a three-level undecimated Haar wavelet transform\cite{unser1995} and correspond to the energies
in the low pass channel and 9 high frequency channels. The
remaining three features are the mean, variance, and entropy of the
pixel intensities in the cytoplasm.

\subsection{Oocyte classification}
\label{sec:classification}

For each oocyte, the extracted feature vector is normalized and fed into
a~binary classifier to produce a~binary label, \textit{viable} or
\textit{nonviable}. Several classifiers were tried with similar results. For the sake of space, only kernel SVM results are reported. 
The RBF SVM kernel with $\gamma = 10^{-2}$ and the cost parameter ${C=1.0}$ were
selected using stratified cross-validation and grid search on a~training dataset of 83
randomly selected oocytes (ROIs). The remaining 20 ROIs were left for testing (Sec.~\ref{sec:classification-results}). This yielded the mean validation accuracy $A_{\text{val}} \approx74\%$.

\section{Results}

First, we evaluate the three stages of the pipeline separately. After that, we examine the significance of the designed features. Finally, we analyze the expert annotations.

\subsection{Oocyte localization evaluation}
\label{sec:localization-results}

Cross-validation was performed to evaluate the oocyte localization
described in Sec.~\ref{sec:localization} using the 34 training images. For
our data, the method worked perfectly. The number of detected oocytes was always
equal to the ground truth and the localization error between the centers of
gravity of the \textit{cytoplasm} class, and the ground truth cytoplasm segmentation
was inferior to 10 pixels (corresponding to approximately 5\% of the oocyte
diameter) in $98\%$ of cases.

\subsection{Oocyte segmentation evaluation}

The five-class oocyte segmentation (Sec.~\ref{sec:segmentation}) was
evaluated using a~10-fold cross-validation on the 103 ROIs, each approximately
centered on one oocyte. 

The Intersection over Union (IoU) metric computed over the folds for the
\textit{cytoplasm}, \textit{zona pellucida}, \textit{polar body}, and \textit{cumulus cells} classes was $95.48\%$,
$89.72\%$, $42.85\%$, and $60.29\%$, respectively.  While \textit{cytoplasm} and
\textit{zona pellucida} are segmented very well, the \textit{polar body} class suffers from false
detections and is more difficult to segment due to its small size.  Although
the \textit{cumulus cells} segmentation performance is also far from perfect, it
is mostly confused with background, which is not critical for our
task (see Fig.~\ref{fig:segmentation}).

\subsection{Oocyte classification evaluation}
\label{sec:classification-results}

The performance of the SVM from Sec.~\ref{sec:classification} to
classify the oocytes as viable or not was evaluated
using the 20 testing ROIs omitted during the training.

\begin{figure}[tb]
\centering
\includegraphics[width=60mm]{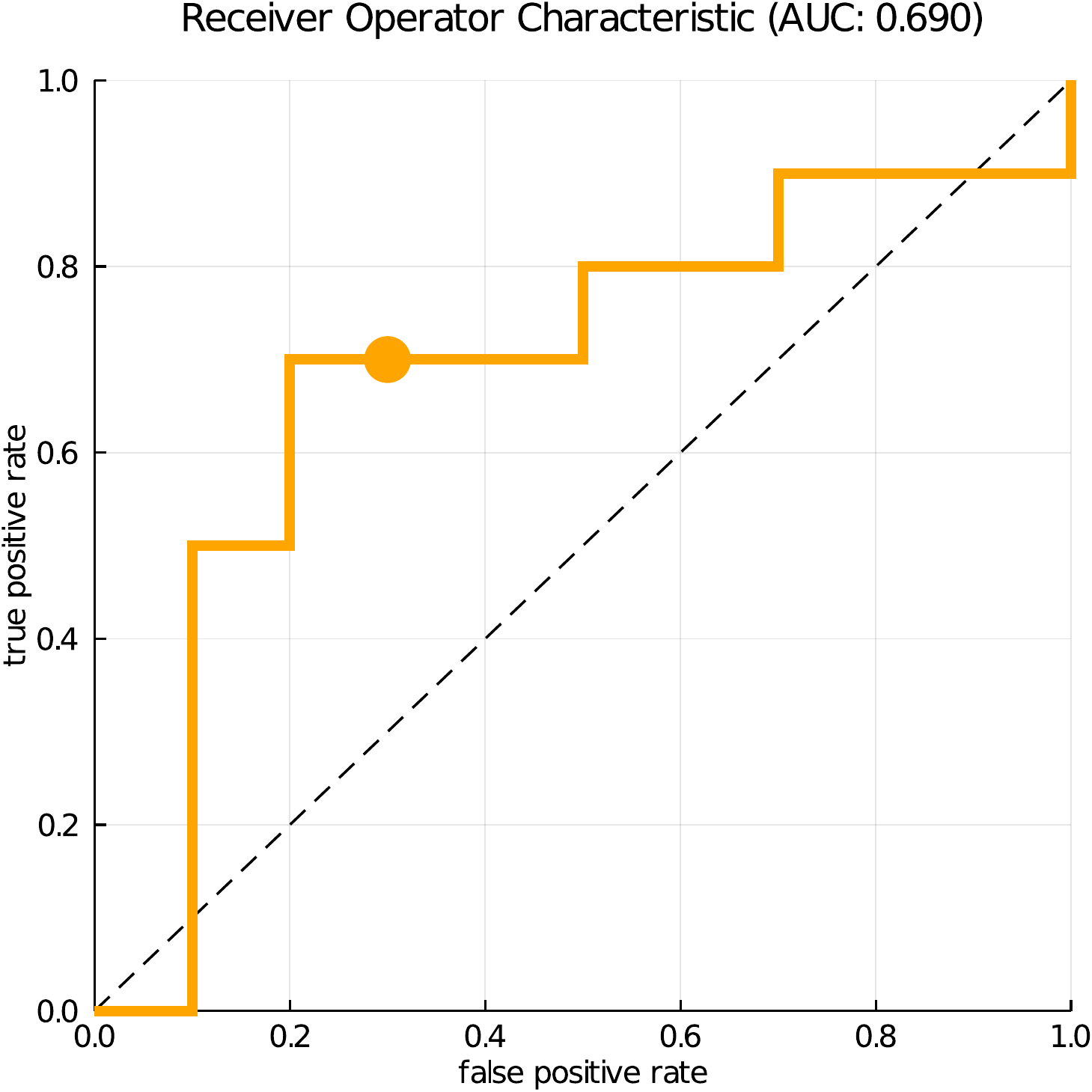}
\caption{ROC curve obtained for the SVM classifier on the testing
  data. The dot represents our operating point.}
\label{fig:svm-roc}
\end{figure}

The classifier obtained the testing accuracy $A_{\text{test}} = 70\%$, sensitivity $Se_{\text{test}}=70\%$, specificity $Sp_{\text{test}}=70\%$, precision $P_{\text{test}}=70\%$, and the area under the ROC curve was $AUC_{\text{test}}=0.69$ (see Fig.~\ref{fig:svm-roc}).

\subsection{Feature significance}

To evaluate the features' significance, we ran a leave-one-out cross-validation on the whole dataset for different subsets of the features. Table~\ref{tab:feature-significance} contains the results for four selected subsets. The most significant feature is the number of polar bodies $n_{pb}$. A viable oocyte contains a single polar body ($n_{pb}=1$). However, in some images, the polar body may be hidden. Both the textural and geometrical features improve the classification accuracy, with a small additional improvement if using both.

\begin{table}[b]
\centering
\caption{Accuracy achieved using four feature subsets. Each subset contains the
number of polar bodies $n_{pb}$. The 13 texture features are denoted as
``texture", and ``geometry" denotes the remaining 10 features. The table shows
the average accuracy computed using the leave-one-out procedure.}
\label{tab:feature-significance}
\begin{tabular}{lrr}
\hline
features & number of features & accuracy \\
\hline
$n_{pb}$            & 1 & $0.515$ \\
$n_{pb}$ + texture  & 14 & $0.738$ \\
$n_{pb}$ + geometry & 11 & $0.767$ \\
all                 & 24 & $0.777$ \\
\hline
\end{tabular}
\end{table}

\subsection{Expert annotation evaluation}
\label{sec:expert-accuracy}

The oocyte annotations created by the embryologist are used to train the classifier. Since these labels are subjective, we evaluate their expected accuracy using the reliably known number of viable oocytes $y_j$ as a reference.

\begin{figure}[t]
\centering
\includegraphics[width=70mm]{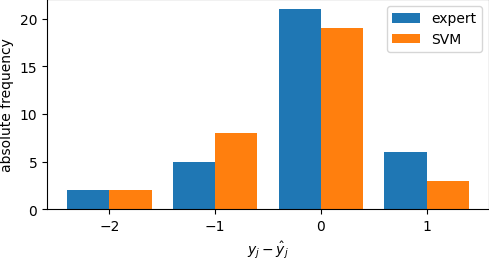}
\caption{Histogram of annotation errors committed by a human expert and an SVM.}
\label{fig:human-error-hist}
\end{figure}

Given image $j$, we denote $n_j$ the number of oocytes and $\hat{y}_j$ the number of oocytes labeled as viable by the expert. The mean absolute error of the expert $\text{MAE} = {\frac{1}{N} \sum_{j=1}^N \lvert y_j - \hat{y}_j \rvert}$, where $N$ is the number of images, was $0.44$ with approximately 3 oocytes per image on the average. Fig.~\ref{fig:human-error-hist} displays a histogram of the differences in the number of viable oocytes per image for the annotations given by the expert and for those predicted by the leave-one-out cross-validation of our automatic method. The histogram reveals only a slight difference between the two annotations. The Kolmogorov-Smirnov statistic computed for the two histograms is $D=0.11$, which suggests no significant difference.

\section{Discussion and conclusions}

In this paper, a proof-of-concept solution was proposed to automatically detect oocytes with good developmental potential and are therefore viable for fertilization.  When interpreting the results, it is important to realize that the expert assessment of the oocyte quality from a~single image is difficult and rather subjective. Our performance (AUC=0.69) is nevertheless comparable to the performance achieved by Manna et al.~\cite{manna2013} (AUC=0.68), who worked with the more reliable information of whether an oocyte leads to birth. The errors in the number of viable oocytes per image are also comparable to that of the expert. We expect our accuracy to improve further when more data is available. This should enable us to employ deep learning techniques instead of hand-crafted features, boosting the performance. We are also working on learning directly from the number of viable oocytes. 

\section{Acknowledgments}
\label{sec:acknowledgments}

The authors acknowledge the support of the OP VVV funded project ``CZ.02.1.01/0.0/0.0/16\_019/0000765 Research Center for Informatics" and the Grant Agency of the Czech Technical University in Prague, grant No. SGS20/170/OHK3/3T/13.

\bibliographystyle{spiebib}
\bibliography{refs}

\end{document}